\documentclass[preprint,superscriptaddress,showpacs,aps]{revtex4}
\usepackage{amsmath}
\usepackage{amssymb}
\usepackage{graphicx}

\newcommand{\dt}{\partial_t}
\newcommand{\vb}{{\bf v}}

\newcommand{\be}{\begin{equation}}
\newcommand{\ee}{\end{equation}}
\newcommand{\bean}{\begin{eqnarray}}
\newcommand{\eean}{\end{eqnarray}}

\begin{document}

\title{Bifurcations and chaos in large Prandtl-number Rayleigh-B\'{e}nard Convection}

\author{Supriyo Paul}
\email{supriyo@iitk.ac.in}
\affiliation{Department of Physics, Indian Institute of Technology, Kanpur~208 016, India}

\author{Pankaj Wahi}
\email{wahi@iitk.ac.in}
\affiliation{Department of Mechanical Engineering, Indian Institute of Technology, Kanpur~208 016, India}

\author{Mahendra K. Verma}
\email{mkv@iitk.ac.in}
\affiliation{Department of Physics, Indian Institute of Technology, Kanpur~208 016, India}

\date{\today}
\begin{abstract}
A low-dimensional model of  large Prandtl-number ($P$) Rayleigh B\'{e}nard
convection is constructed using some of the important modes of
pseudospectral direct numerical simulations.  A detailed bifurcation
analysis of the low-dimensional model for $P=6.8$ and aspect ratio of $2\sqrt{2}$
reveals a rich instability and
chaos picture: steady rolls, time-periodicity, quasiperiodicity,
phase locking, chaos, and crisis.  Bifurcation analysis also reveals
multiple co-existing attractors, and a window with time-periodicity after chaos. The results of the low-dimensional model matches quite closely with some of the
past simulations and experimental results where they observe chaos in RBC
through quasiperiodicity and phase locking.

\end{abstract}


\maketitle

\section{Introduction}

Rayleigh B\'{e}nard convection (RBC) exhibits a host of complex
phenomena.  Instabilities, patterns, and chaos are observed near the
onset of convection, while turbulence is seen far beyond the
onset~\cite{rbc_intro}.  Bifurcation diagrams are often used to
study instabilities and chaos.  In this paper we report a
bifurcation analysis for  large-Prandtl number RBC flow obtained using a low-dimensional model.

The two most important parameters for RBC are the Rayleigh number $R$
(ratio of buoyancy and dissipative force) and the Prandtl number $P$
(ratio of viscosity and thermal diffusivity).  Convection starts at
the critical Rayleigh number $R_c$, which is independent of $P$. For
non-zero Prandtl number, the primary instability  at the onset of convection is always in the form of two-dimensional (2D) straight rolls~\cite{schlutter:jfm_1965}. These rolls become
unstable through secondary instabilities, and they bifurcates into a
sequence of dynamic
patterns~\cite{busse:jfm_1971,busse:rpp_1978,busse_clever:jfm_1979}.
Some examples of the resulting patterns are squares and hexagons,
asymmetric patterns, oscillating patterns, relaxation oscillations
of rolls and squares, etc.~\cite{patterns}.

The sequence of instabilities and onset of chaos are  quite
different  for low-Prandtl number (low-P)
convection~\cite{Libchaber} and large-Prandtl number (large-P)
convection~\cite{Giglio:1981,berge:jpl_1980,gollub:jfm_1980,stavans:prl_1985}.
For low-P convection, the 2D rolls become unstable close to
the onset and wavy rolls are generated through secondary
instabilities, thus making the flow three-dimensional (3D). These
bifurcations and route to chaos for low-P and zero-Prandtl (zero-P)
number convection have been studied extensively (see
\cite{rbc_intro,low_P,Pal:zeroP} and references therein).   The scenario
however is quite different for large-P convection. The secondary
instabilities are delayed here, and the 2D rolls continue to be
solutions till larger Rayleigh numbers. It has been reported that 2D
convection results have significant similarities with
3D results for large-P
convection~\cite{vincent:pre_2000, schmalzl:epl_2004}.  We exploit
this observation to analyze bifurcation and chaos for large-P
convection using a low-dimensional model containing only 2D modes.
A major advantage of this simplification is that the number
of modes required for 2D convection is much fewer than 3D convection,
thus enabling the bifurcation analysis.

Krishnamurti~\cite{krishnamurti:jfm_1970} performed extensive
convection experiments on mercury ($P \approx 0.02$),  air ($P
\approx 0.7$), water  ($P \approx 6.8$), freon 113  ($P \approx
7$), and silicon oil ($P \sim 100$). She studied
transition from 2D convection to 3D convection and subsequent
generation of oscillatory, chaotic, and turbulent convection. Busse
and Whitehead~\cite{busse:jfm_1971} also reported `zigzag
instability' and `cross-roll instability' in an experiment on
silicon oil.

Libchaber {\it et al.}~\cite{Libchaber} studied the routes to chaos
for RBC in mercury (a low-P fluid) as a function of $R$ and the
applied mean magnetic field; at different values of the parameters
they observed chaos through various routes: period doubling,
quasiperiodic, and soft instability.  Gollub and
Benson~\cite{gollub:jfm_1980} studied route to chaos in RBC of water
at two different Prandtl numbers, $P=2.5$ and $5$, for different
aspect ratios $\Gamma$ and observed very rich behaviour.  For $P=5$
and  $\Gamma=3.5$ they observed steady rolls till $r=R/R_c = 27.2$
($r$ is called `reduced Rayleigh number'), at which point periodic
flow starts.  At $r=32$, a second frequency appears in the system
and the flow becomes quasiperiodic.  Phase locking occurs at
$r=44.4$ that finally leads to chaos at $r=46.0$.  Gollub and Benson
also observed period doubling route to chaos for $P=2.5$ and
$\Gamma=3.5$, and quasiperiodic route to chaos with the third
frequency for $P=5$ and $\Gamma=2.4$. They also observed
intermittency in their system.

In another experiment, Maurer and Libchaber~\cite{Maurer:1979}
observed frequency locking and subsequent generation of chaos in liquid helium
as a result of generation of new frequency modes.  Giglio {\it et
al.}~\cite{Giglio:1981} observed period-doubling route to chaos in
their convective experiment on water.  Berg\'{e} et
al.~\cite{berge:jpl_1980} found intermittency in RBC of silicon oil.
Ciliberto and Rubio~\cite{Ciliberto:physcripta_1987} reported
localized oscillations and travelling waves in RBC. Morris {\it et
al.}~\cite{morris:prl_1993} discovered spatio-temporal chaos in RBC
of silicon oil.  These results indicate complex nonlinear
dynamics including chaos in RBC.

Direct numerical simulation  (DNS) of 2D and 3D RBC have been used
to study various convective states including turbulence.  Curry {\it
et al.}~\cite{curry:jfm_1984} performed detailed 3D DNS for $P=10$
under free-slip  boundary conditions; they reported steady convection till
$r\approx 40$, after which single frequency oscillations are
observed till $r\approx 45$. Subsequently they observed
quasiperiodicity ($r \approx 45-55$), phase locking ($r\approx
55-65$), and chaos ($r > 65$).   Yahata~\cite{Yahata:2000} performed
DNS using finite difference scheme (MAC method) on no-slip boundary
condition for $P=5$, $\Gamma_x=3.5$ and $\Gamma_y = 2$. The results
show a series of bifurcations from monoperiodic $\rightarrow$
biperiodic $\rightarrow$ frequency-locked state $\rightarrow$
chaotic state.  Mukutmoni and Yang~\cite{Mukutmoni:1993} reported a numerical study
of  RBC in a rectangular enclosure with insulated sidewalls. They
observed a period-2 response after a periodic solution but the route
to chaos is through quasiperiodicity.
However, on imposing symmetry of the velocity and temperature field
about the mid-planes, they observed a period-doubling route to
chaos. They have also reported periodic solutions after chaos.

In two dimensions, Moore and Weiss~\cite{moore_weiss:jfm_1973} simulated $P=6.8$ RBC using a
spectral method with free-slip boundary conditions; they studied heat transport as a function of
the Rayleigh number.  McLaughlin and
Orszag~\cite{mclaughlin_orszag:jfm_1982} considered RBC in air ($P =
0.71$) with no-slip boundary conditions; they obtained periodic,
quasiperiodic, and chaotic states for Rayleigh numbers between 6500
and 25000.  Curry {\it et al.}~\cite{curry:jfm_1984} performed detailed
DNS for $P=6.8$ fluid in 2D and observed oscillations
with a single frequency at $r \approx 50$, and
with two frequencies at $r \approx 290$. They also
reported weak chaos beyond $r = 290$, and a  periodic
solution after $r \approx 800$.   Goldhirsch {\em et
al.}~\cite{goldhirsch:jfm_1989} also simulated 2D RBC and observed complex behaviour.

Recently, Paul {\em et al.}~\cite{paul:arxiv_2009} performed 2D DNS
for free-slip boundary condition for a large range of Rayleigh numbers and
obtained steady convection ($r = 0-80$), periodicity ($r = 80-660$),
quasiperiodicity ($r = 660-770$), and chaos ($r = 770-890$).  They
also observed periodic and steady convection beyond the chaotic
state.   Paul {\em et al.}~\cite{paul:arxiv_2009}'s results are in general
agreement with those of Curry {\it et al.}~\cite{curry:jfm_1984} and
Goldhirsch {\em et al.}~\cite{goldhirsch:jfm_1989} with some
difference in the Rayleigh numbers.  One noticeable difference
between 2D RBC and 3D RBC is that the secondary
instabilities in 2D RBC occur at significantly higher values of $r$
than those in 3D RBC.

The origin of various convective patterns and chaos in DNS is not
apparent due to various interactions among the large number of
modes. Rather, they have been analyzed using low-dimensional models
of RBC.  Curry~\cite{Curry:PRL} constructed a 14-mode model of RBC
with a small amplitude periodic modulations in the heat equation. He
observed chaos for $P=10$.  The low-dimensional model of Curry shows
features similar to the experiments of Gollub and
Benson~\cite{gollub:jfm_1980} namely periodicity, quasiperiodicity,
and chaos. Curry~\cite{Curry:CommMath} also studied the 14-mode
model without any modulation, and compared its results
with those from the Lorenz model.

Yahata~\cite{Yahata:1982} studied transition to chaos in RBC using a
48-mode system of equations under no-slip boundary conditions.  For
$P=5$, $\Gamma_x = 2$ and $\Gamma_y=3.5$, he obtained periodic
$\rightarrow$ quasiperiodic motion with two fundamental frequencies
$\rightarrow$ quasiperiodicity with three frequencies $\rightarrow$
chaos.  The whole sequence of bifurcations occur in the range of
$r=39.77$ to $41.04$.  Yahata~\cite{Yahata:1983} continued the above
analysis for $P=2.5$ with the same aspect ratio and reported
period-doubling route to chaos for $r$ in the range of $24.46$ to
$29.35$.

In the above work, Curry~\cite{Curry:PRL} and
Yahata~\cite{Yahata:1982,Yahata:1983} numerically integrate  the
low-dimensional models for certain $r$ values and observe various
patterns.  Some of the shortcomings of this method are that it could
miss the behaviour in narrow windows, and the bifurcation points cannot be precisely located.  In the present paper we numerically time advance the fixed
points, limit cycles, and chaotic attractors, which provides a much
more detailed bifurcation picture than that obtained by studying
patterns at selected $r$ values.  The approach in the present paper
is very similar to a recent work by Pal {\em et
al.}~\cite{Pal:zeroP}  where a detailed bifurcation
diagram was constructed for zero-P convection using a 13-mode system; this system was obtained using the energetic modes of DNS. Note  that in
large-P convection chaos sets in at a much larger Rayleigh number
than in low-P and zero-P convection.  As a consequence, large-P
convective flows contain a large number of energetic modes at the
onset of chaos. A bifurcation analysis of these large number of
modes is very difficult and impractical.  To circumvent this
difficulty we study 2D convection, which is not a serious limitation
for large-P convection since 2D and 3D convection have significant
similarities~\cite{vincent:pre_2000, schmalzl:epl_2004}.

We perform our low-dimensional analysis for $P=6.8$,
which is the Prandtl number of water, a representative of large-P
fluid.  A significant advantage of choosing this Prandtl number is
that a large number of DNS and experiments have been performed for
water.  We find that optimal number of real Fourier modes of our
low-dimensional model is 30.  The convective patterns--steady,
periodic, quasiperiodic, and chaotic rolls--observed in DNS and experiments are
captured in our low-dimensional model.

The outline of the paper is as follows.  In Section II we describe
the governing equations and the low-dimensional model. 
Details of the various convective regimes, bifurcation
scenario and route to chaos associated with this model is presented
in section III. The last section contains discussions and  conclusions.

\section{The low-dimensional model and its numerical simulation}

In RBC,  a thin layer of fluid is confined between two thermally
conducting horizontal plates that are separated by a distance $d$.
The fluid has kinematic viscosity $\nu$, thermal diffusivity
$\kappa$, and thermal expansion coefficient  $\alpha$. An adverse
temperature gradient $\beta=\Delta{T}/d$ is imposed across the fluid
layer, where $\Delta{T}$ is the temperature difference across the
layer. We assume Boussinesq approximation for the fluid
\cite{rbc_intro}.  The relevant hydrodynamic equations are
nondimensionalized using the length scale $d$,  the large-scale velocity scale
$\sqrt{\alpha\beta g d^{2}}$, and the temperature scale $\Delta{T}$
to yield \cite{rbc_intro} \bean
\dt\vb + (\vb\cdot\nabla)\vb &=& -\nabla{p}+\theta\hat{z}+\sqrt{\frac{P}{R}}\nabla^2 \vb, \label{NS_eqn}\\
\dt\theta + (\vb\cdot\nabla)\theta &=& v_3 +\frac{1}{\sqrt{PR}} \nabla^2 \theta, \label{heat_eqn} \\
\nabla \cdot \vb & = & 0, \label{continuity} \eean
where
$\vb=(v_1,v_2,v_3)$  is the velocity fluctuation, $\theta$ is the
perturbations in the  temperature field from the steady conduction
state, $R=\alpha g \beta d^4/\nu \kappa$ is the Rayleigh number,
$P=\nu/\kappa$ is the Prandtl number, $g$ is the acceleration due to gravity, and $\hat{z}$ is the buoyancy
direction.

The above equations are often solved using direct numerical
simulation (DNS). One of the popular numerical technique is
pseudospectral method in which the velocity and the temperature
fields are expanded in the Fourier/Chebyshev basis.   These numerical
simulations have been able to reproduce various patterns, chaos, and
turbulence observed in experiments.  The convection simulations are
however very expensive in terms of computer time and memory.  Also a
large number of modes present in DNS obscures the internal dynamics.
A popular method to analyze such systems is a bifurcation analysis of
appropriate low-dimensional systems.  Using this technique we can
study the origin of various patterns and chaos in RBC.   For our
low-dimensional model we choose fourteen complex modes and two real
modes that represent the large-scale flow structures. Expansion of
the vertical velocity field $v_{3}$ and the temperature field
$\theta$ using these modes yields
\bean
v_{3}(x,z,t) &=& W_{101}(t)\exp(ik_{c}x)\sin(\pi z)  + W_{103}(t)\exp(ik_{c}x)\sin(3\pi z) \nonumber \\
&+& W_{105}(t)\exp(ik_{c}x)\sin(5\pi z) + W_{202}(t)\exp(2ik_{c}x)\sin(2\pi z) \nonumber \\
&+& W_{301}(t)\exp(3ik_{c}x)\sin(\pi z)  + W_{303}(t)\exp(3ik_{c}x)\sin(3\pi z) \nonumber \\
            &+& W_{501}(t)\exp(5ik_{c}x)\sin(\pi z) + c.c. , \nonumber \\
v_{2}(x,z,t) & = & 0, \nonumber \\
\theta(x,z,t) &=&   \theta_{101}(t)\exp(ik_{c}x)\sin(\pi z) + \theta_{103}(t)\exp(ik_{c}x)\sin(3\pi z)\nonumber \\
      &+& \theta_{105}(t)\exp(ik_{c}x)\sin(5\pi z) + \theta_{202}(t)\exp(2ik_{c}x)\sin(2\pi z) \nonumber \\
      &+& \theta_{301}(t)\exp(3ik_{c}x)\sin(\pi z)  + \theta_{303}(t)\exp(3ik_{c}x)\sin(3\pi z) \nonumber \\
        &+& \theta_{501}(t)\exp(5ik_{c}x)\sin(\pi z) + c.c. \nonumber \\
        & + & \theta_{002}(t)\sin(2 \pi z) + \theta_{004}(t)\sin(4 \pi z)
            \label{model_expansion}
\eean
where {\em c.c.} stands for the complex conjugate, and the three subscripts denote the Fourier wavenumber indices along $x$, $y$, and $z$ directions respectively. These modes correspond to the free-slip boundary condition for the velocity modes.  Note that
$v_1(x,z)$  can be computed using the incompressibility condition
$\nabla \cdot {\bf v} = 0$.  We choose $k_c = \pi/\sqrt{2}$, hence the aspect ratio of our model is $2\sqrt{2}$.  The Galerkin projection of
Eqs.~(\ref{NS_eqn}-\ref{continuity}) on these modes yields a set of
thirty coupled ordinary differential equations (ODEs) for the real
and imaginary parts of the Fourier modes. These thirty nonlinear
ODEs comprise our low-dimensional model.

The above modes represent two-dimensional rolls.  It has been
reported earlier that the 2D and 3D convection have significant
similarity for large-P flows~\cite{schmalzl:epl_2004}.  Therefore we
expect our low-dimensional model to capture the dynamics of large-P
convection. Our model with 2D rolls has 30 modes while a full 3D
low-dimensional model would have many more modes that would make the
bifurcation analysis of the model very difficult.  Note that
three-dimensional patterns like squares are not accessible to our
model. However quasiperiodicity and the origin of chaos are expected
to be common for both 2D and 3D convection for large Prandtl number
flows.

Our low-dimensional model is quite similar to that of
Curry~\cite{Curry:PRL}. A major difference is that in our model all
the modes except $\theta_{002}$ and $\theta_{004}$ are complex in
contrast to Curry's model in which they are all real.  Also, we
keep the mode $(105)$, whereas Curry keeps $(204)$. Several of the
patterns and chaos reported in the experiments of Gollub and Benson~\cite{gollub:jfm_1980} have been observed by Curry
when he includes small amplitude modulation.  We do not require any
modulation or any additional forcing (other than buoyancy) in our
model to produce these patterns and chaos.
Yahata~\cite{Yahata:1982,Yahata:1983} studied RBC under no-slip
boundary condition by expanding the velocity and temperature fields
using mixed basis functions (Chebyshev along the buoyancy direction
and Fourier along the horizontal directions) and observed similar behaviour.  Surprisingly the
patterns and chaos reported for the no-slip and the free-slip
boundary conditions are quite similar.

\begin{figure}[t]
\begin{center}
\includegraphics[height=!,width=14cm]{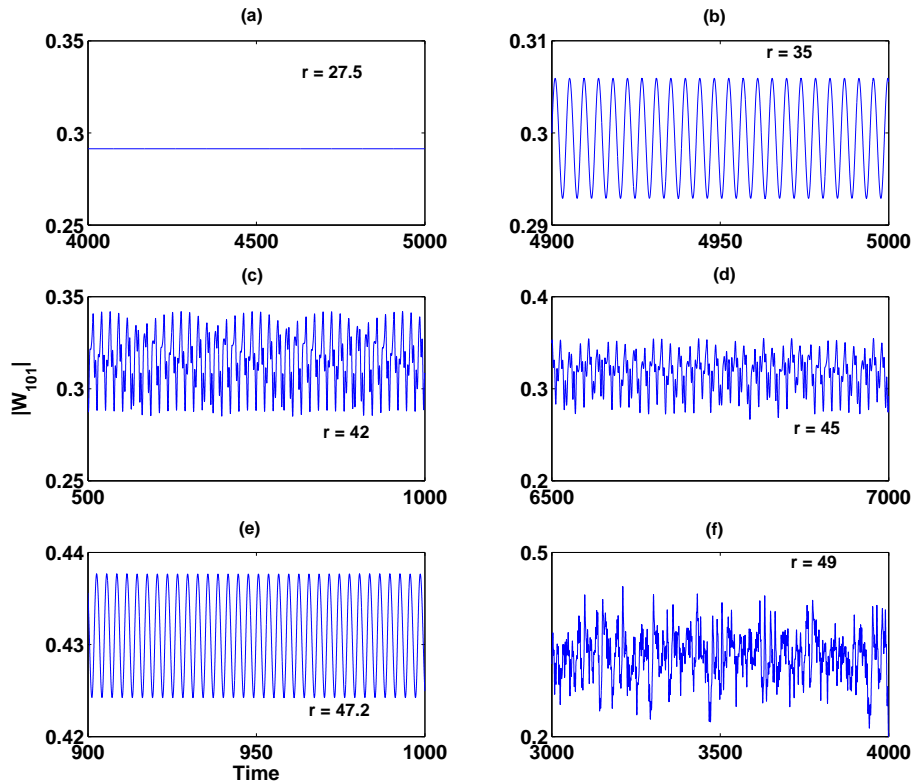}
\end{center}
\caption{Time series of the amplitude of the complex mode $W_{101}$
generated by the low-dimensional model at various representative reduced Rayleigh
numbers ($r$). We observe (a) steady convection ($r=27.5$), (b) time-periodic
convection ($r=35$), (c) quasi-periodicity ($r=42$) and (d) chaos ($r=45$).
Subsequently, (e) a window of time-periodic state ($r=47.2$) followed by (f) a chaotic state ($r=49$)  is observed.} \label{fig:timeseries_low_dim_w11}
\end{figure}

We numerically solve the low-dimensional model using random initial
conditions. In our low-dimensional model, we observe various
patterns: steady convection, periodicity, quasiperiodicity, and
chaos at different values of Rayleigh numbers (see
Fig.~\ref{fig:timeseries_low_dim_w11}).  Figure~\ref{fig:timeseries_low_dim_w11}) also shows
that the system becomes periodic after chaos, and then it becomes
chaotic again.
\begin{figure}[t]
\begin{center}
\includegraphics[height=!,width=14cm]{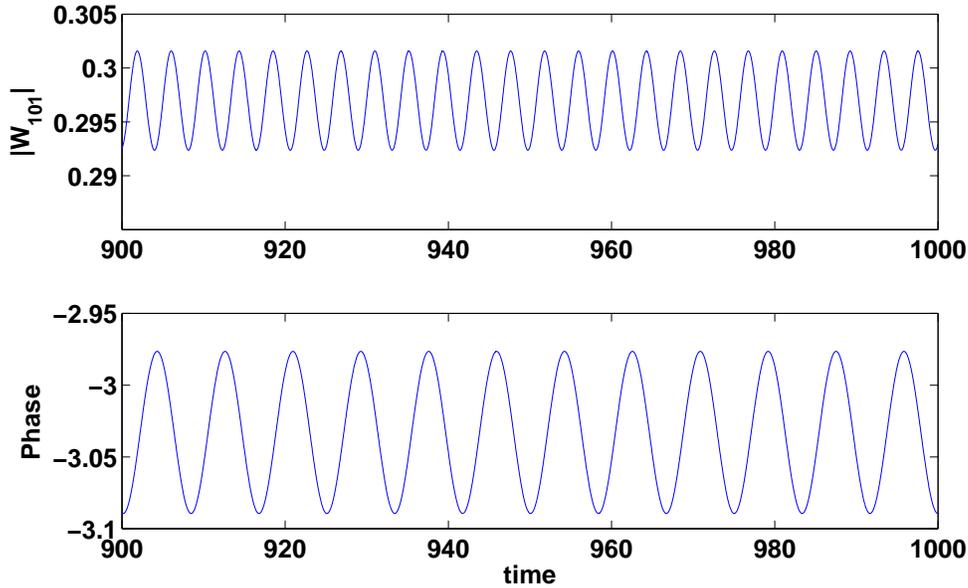}
\end{center}
\caption{Time-series of amplitude and phase of the complex $W_{101}$
mode at $r=30$. The amplitude oscillates at double the frequency of
its phase.} \label{r30_model}
\end{figure}

A curious feature of our model is that for the periodic solutions
($r=27.6-40.3$), the frequency of the Fourier amplitude is twice that
of its phase (see Fig.~\ref{r30_model}).  This feature can also be
understood using the bifurcation analysis that will be discussed
below.

The origin of the observed patterns can be studied more rigorously
using the bifurcation analysis which is the subject of the next
section.

\section{Bifurcation analysis of the low-dimensional model}

\begin{figure}[t]
\begin{center}
\includegraphics[height=!,width=14cm]{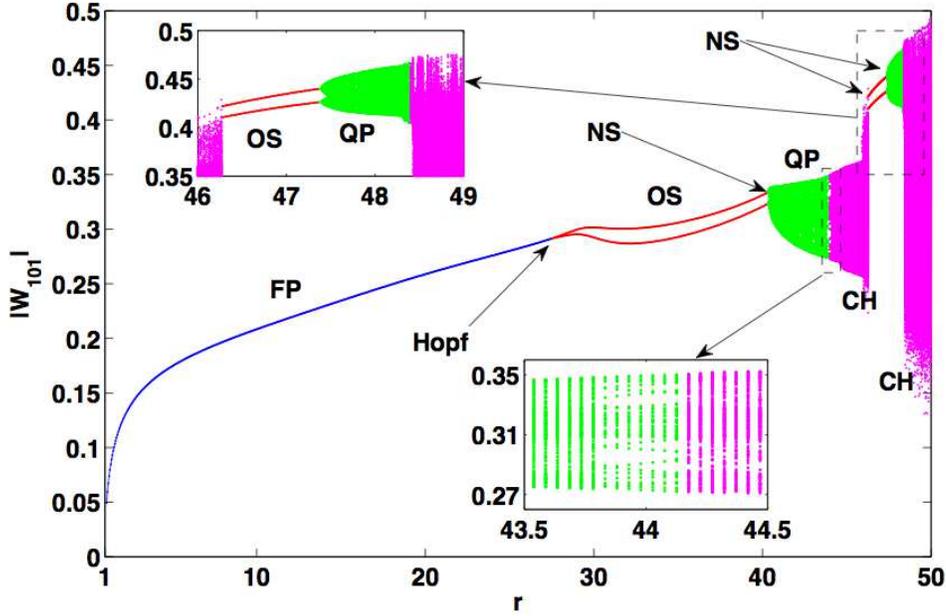}
\end{center}
\caption{Bifurcation diagram of the low-dimensional model
representing large-Prandtl number  RBC. `FP' (blue curve) is the steady roll, `OS' (red
curve) is the time-periodic roll, `QP' (green patch) is the quasi-periodic roll, and `CH' (pink patch) is the chaotic state.  `NS' indicates  the Neimark-Sacker bifurcation point.  A window of periodic and quasiperiodic states is observed in the band of $r=46.2-48.4$. }
\label{bifurcation_model_a}
\end{figure}

In Fig.~\ref{bifurcation_model_a} we present a bifurcation diagram
obtained by numerical integration of the low-dimensional model for $P=6.8$ and the aspect ratio of $2\sqrt{2}$. The reduced Rayleigh number $r$ is the bifurcation parameter in our
analysis. The unstable solutions have not been shown in
Fig.~\ref{bifurcation_model_a}.  To generate this bifurcation
diagram, numerical simulations have been performed with a fixed
initial condition till $t=20000$ (large-scale eddy turnover time). Transients till $t=5000$ are
eliminated and the extremum values of  $|W_{101}|$ are plotted
for later time. The stability and bifurcations of the
steady states in this numerically generated bifurcation diagram are
complimented by an eigenvalue analysis of the jacobian evaluated at
the fixed points, and the eigenvalue of the associated Floquet matrix
for the limit cycles. For this complimentary analysis, a fixed point
is obtained numerically using the Newton-Raphson method for a given
$r$, and the branches of the fixed points are subsequently obtained using a
fixed arc-length based continuation scheme (similar to the analysis
in Pal {\em et al.}~\cite{Pal:zeroP}).

For $r<1$, there is no convection in the system and heat is
transported solely by conduction. The conduction state corresponds
to the trivial  fixed point of our system.  At $r=1$, the system undergoes a pitch-fork bifurcation
and time-independent convection states are born as non-trivial fixed
points. Note that there are two nonzero solutions near the onset of
convection (e.g, $\pm W_{101}$).  In Fig.~\ref{bifurcation_model_a}
we plot $|W_{101}|$ vs.\ $r$.  The new stable roll solutions (blue
curve labeled `FP')  remain stable till $r \approx 27.6$.

The branch of fixed points corresponding to the steady convective
rolls undergoes a supercritical Hopf bifurcation at $r=27.6$. As a
consequence, the time-independent steady state solution  becomes unstable and
a new stable time-periodic state (limit cycle) is born.  The limit cycle solution is shown as a red line with the label `OS'  in the bifurcation diagram (Fig.~\ref{bifurcation_model_a}). The two lines of `OS'
state designate the maxima and minima of $|W_{101}|$ respectively.
The time variation of the modes is however more complex.  As shown
in Fig.~\ref{r30_model}, the amplitude of the mode $W_{101}$ vary
with frequency twice that of its phase.  This phenomena can be
understood as follows. At the Hopf bifurcation point, the
eigenvectors associated with the pair of purely imaginary
eigenvalues $\pm i \omega$ have components only along the imaginary
part of the Fourier modes (e.g., $\Im({W_{101}}$)). Hence,
$\Im({W_{101}})$ oscillates with the frequency $\omega$ corresponding
to the Hopf point.  The real parts of the
Fourier modes are generated purely due to the quadratic
nonlinearities involving products of two imaginary parts of the
modes; hence $2 \omega$ (superharmonic) is the leading frequency of
the real parts and the amplitudes of modes.

We determine the stability of the above time-periodic state using
the Floquet theory. We numerically construct the fundamental
(Floquet) matrix associated with the time-periodic state and compute
its eigenvalues (called `Floquet multipliers').  All the
Floquet multipliers  for a stable limit cycle have magnitude less than one. For
an unstable limit cycle, at least one of them has a magnitude greater than one.  When
the Floquet multipliers cross along the positive real axis, new
limit cycles may appear or disappear (`pitchfork' or `turning point'
bifurcation). If they cross the negative real axis, the frequency of
the limit cycle doubles in a `period-doubling' bifurcation. However,
a pair of complex-conjugate multipliers may also cross the unit
circle in the complex plane, wherein another frequency is generated.
This bifurcation is known as `Niemark-Sacker' (NS) or a secondary
Hopf bifurcation.

\begin{figure}[t]
\begin{center}
\includegraphics[height=!,width=14cm]{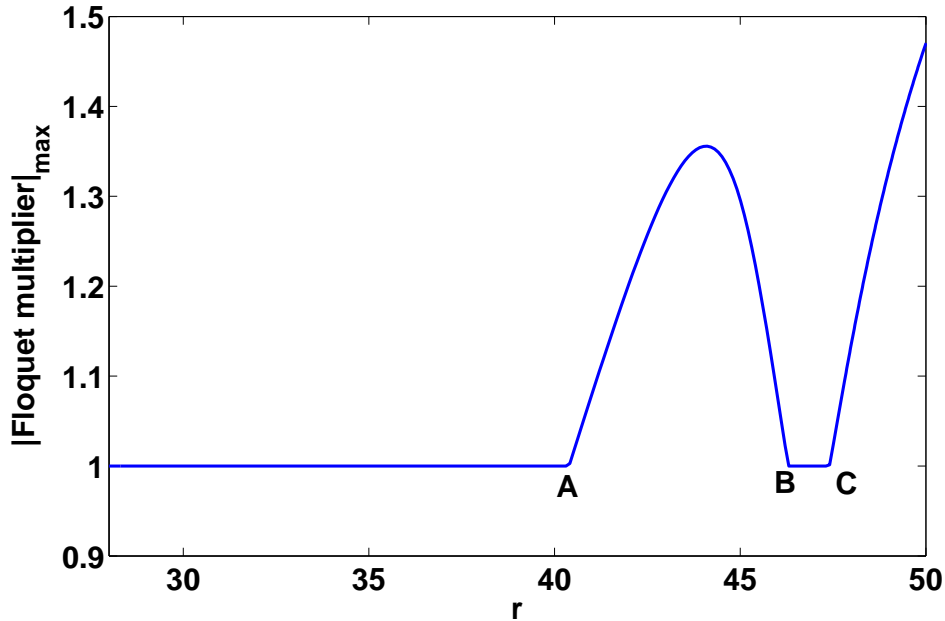}
\end{center}
\caption{The largest Floquet multiplier of the low-dimensional
system as a function of $r$. The largest Floquet multiplier is greater than 1 beyond $r=40.3$ (point `A') except in the `BC' window in which we observe periodic states. }
 \label{floquet}
\end{figure}

\begin{figure}[t]
\begin{center}
\includegraphics[height=!,width=18cm]{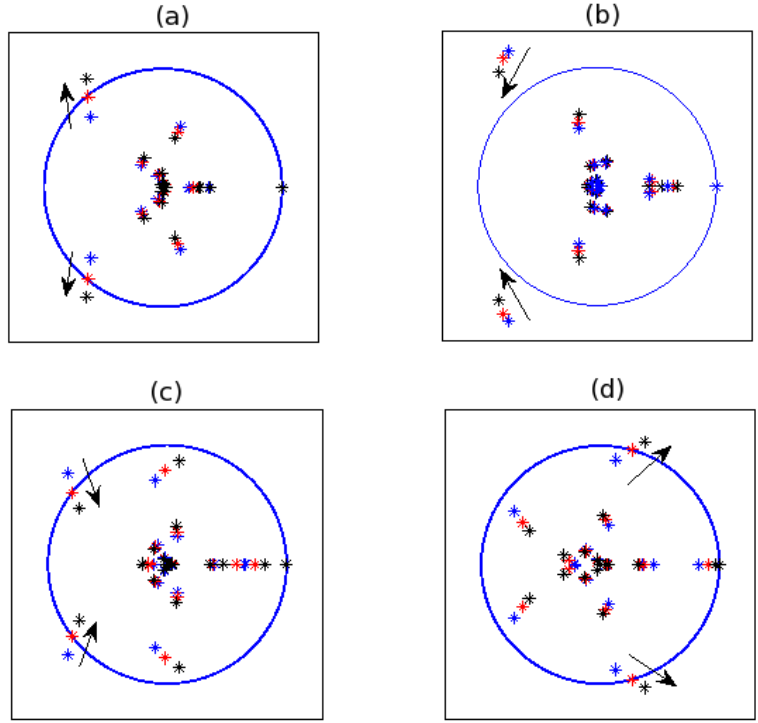}
\end{center}
\caption{Representation of the Floquet multiplier (FM) for various scenarios. Blue points indicate initial stage, red points indicate intermediate stage, and grey points indicate the final stage of the movement of the FM.
(a) The FM crosses the unit circle through NS bifurcation creating a quasiperiodic state. (b) The motion of the FM during the phase-locked regime. Note that the largest FM remain outside the unit circle in this regime. (c) The largest FM  moves into the unit circle resulting in a periodic solution.  (d) The largest FM crosses the unit circle again creating a quasiperiodic solution.}
\label{floquet_plane}
\end{figure}

Fig.~\ref{floquet} illustrates the magnitude of the largest Floquet
multiplier as a function of the reduced Rayleigh number $r$, while
Fig.~\ref{floquet_plane} shows the movement of the Floquet
multipliers around several values of $r$. For these
calculations, we proceed as follows. The limit cycle along with its
time-period is obtained as a fixed point of an appropriately
defined map (described in appendix \ref{Poincare_map}) using the
Newton-Raphson method for a given $r$. The branch of limit cycles is
subsequently computed using a fixed arc-length based continuation
scheme. The fundamental matrix and the eigenvalues associated with
the limit cycles for each value of $r$ are then evaluated numerically.

\begin{figure}[t]
\begin{center}
\includegraphics[height=!,width=14cm]{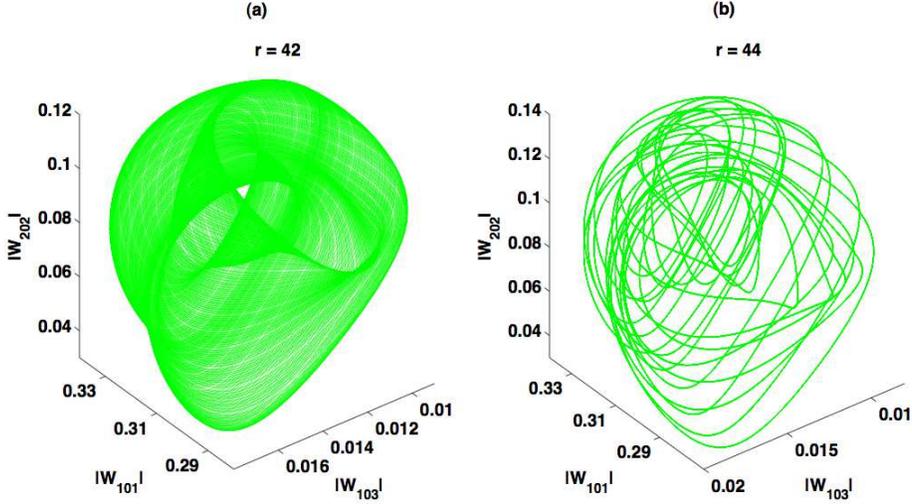}
\end{center}
\caption{A three-dimensional phase space projection of the phase space
trajectories.  (a) At $r=42$ the phase space trajectories  fill the torus (quasiperiodic). (b) At $r=44$ the system is in a phase-locked state, and the phase space trajectory does not fill the torus (limit cycle).} \label{torus}
\end{figure}

\begin{figure}[t]
\begin{center}
\includegraphics[height=!,width=14cm]{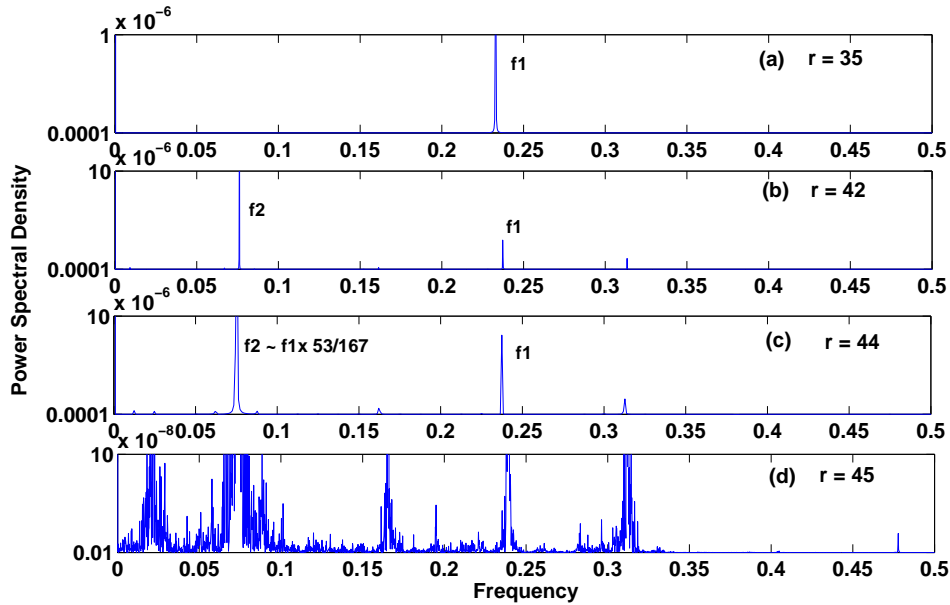}
\end{center}
\caption{The power spectral density of the mode $W_{101}$ for various dynamical states: (a) periodic, (b) quasiperiodic ($f_1/f_2 \approx 3.099$), (c) phase locked ($f_1/f_2 \approx 3.152 \sim 167/53$), (d) chaos.}
\label{fig:powerspectra}
\end{figure}

Our computations reveal that the largest Floquet multiplier has magnitude less than one up to $r
\approx 40.3$ (till `A' in Fig.~\ref{floquet}), hence the limit
cycle (periodic orbit) remains stable till $r \approx 40.3$.  The
system undergoes a Niemark-Sacker (NS)  near $r \approx 40.3$ as illustrated in Fig.~\ref{floquet_plane}(a). As a  result of NS bifurcation,  a second frequency (incommensurate with
the first one) is generated, and the phase space trajectories show a
transition from periodic orbits to quasiperiodic orbits.  The
quasiperiodic  state is shown as a green patch labeled `QP' in the
bifurcation diagram (Fig.~\ref{bifurcation_model_a}). The phase
space  trajectories in this regime lie on a torus as illustrated in
Fig.~\ref{torus}(a) for $r=42$.  The power spectral density of the
mode $|W_{101}|$ shown in  Fig.~\ref{fig:powerspectra}(b) have  two leading frequencies whose
frequency ratio is is approximately 3.099.  To identify the leading frequency, the power spectral density for the periodic state is shown in Fig.~\ref{fig:powerspectra}(a).
\begin{figure}[t]
\begin{center}
\includegraphics[height=!,width=14cm]{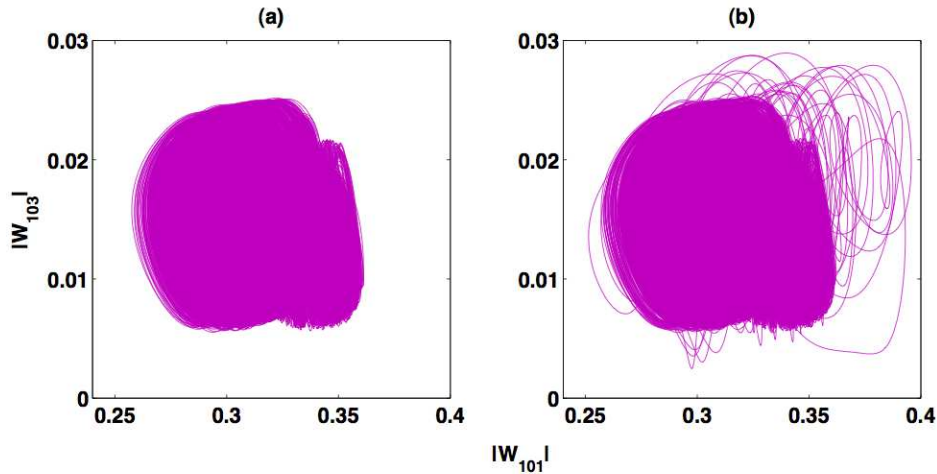}
\end{center}
\caption{The phase space projection on the $|W_{101}|-|W_{103}|$
plane of the chaotic attractor at (a) $r=45.8$ and (b) $45.9$. The chaotic
attractor shows a sudden increase in size at $r=45.9$.}
\label{fig:crisis}
\end{figure}
\begin{figure}[t]
\begin{center}
\includegraphics[height=!,width=14cm]{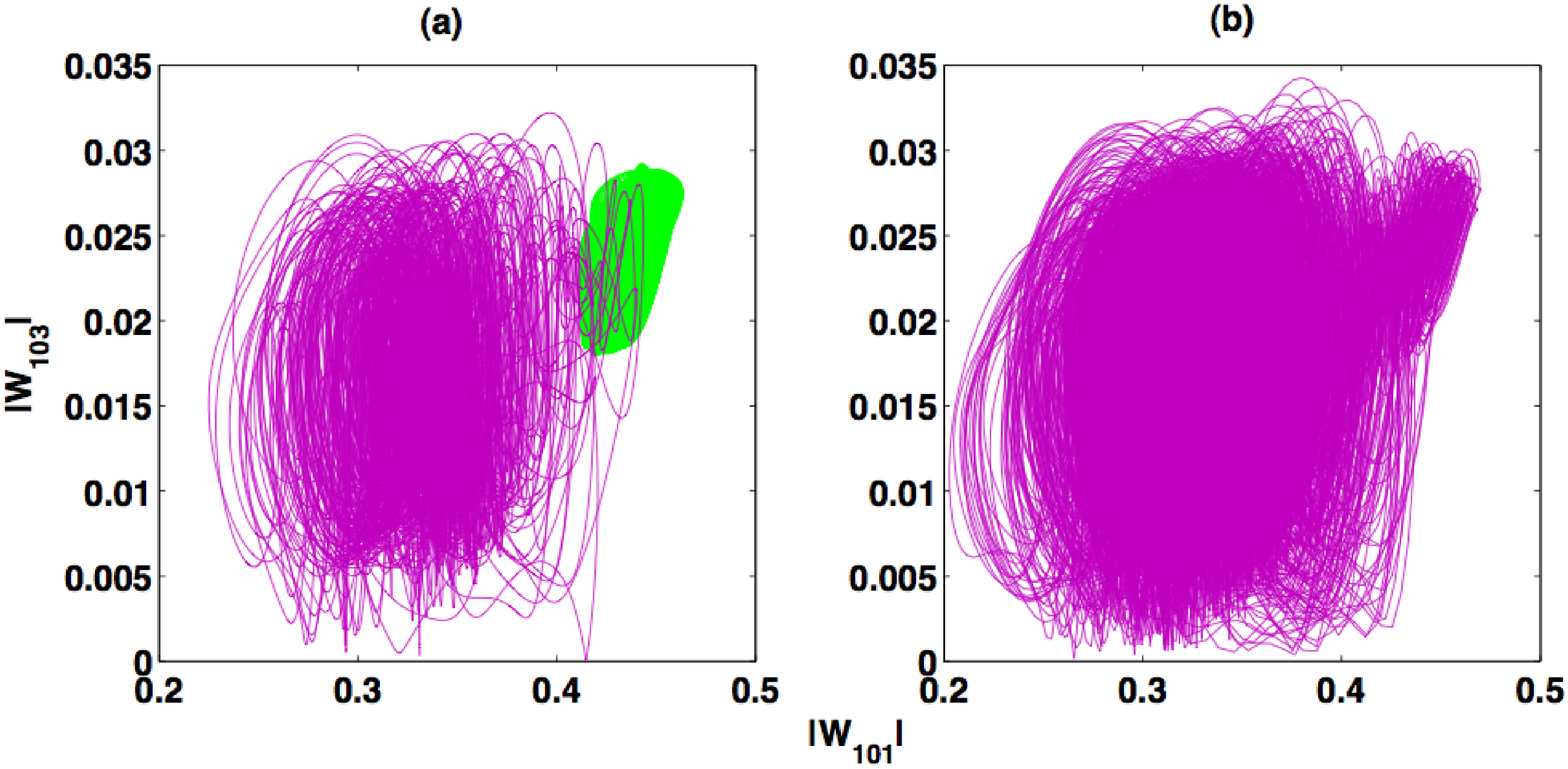}
\end{center}
\caption{The phase space projection on the $|W_{101}|-|W_{103}|$
plane at (a) $r=48.3$ and (b) $48.5$. (a) At $r=48.3$,  two attractors, a chaotic attractor (pink patch) and a quasi-periodic attractor (green patch) coexist. These two attractors are accessible through two different initial conditions. (b) At $r=48.5$ the two attractor merges to generate a larger chaotic attractor.}
\label{fig:crisis2}
\end{figure}

As the bifurcation parameter $r$ is increased further, the two
frequencies get phase locked to yield a periodic solution. This
occurs in a narrow window  in
Fig.~\ref{bifurcation_model_a} and its lower inset. The phase space
trajectories still lie on a torus but they form a periodic orbit as
shown in Fig.~\ref{torus}(b) for $r=44$.  The time-period of this
periodic orbit is approximately 150 non-dimensional time units, and
the ratio of the two leading frequencies $f_1/f_2$ is approximately 3.152 ($\sim 167/53$). In Fig.~\ref{floquet_plane}(b) we show
the movement of the Floquet multipliers in the phase locked region.
During this movement of the largest Floquet multipliers the two frequencies get phase locked.
The power spectrum of the mode $W_{101}$ for the phase-locked state is
illustrated in Fig.~\ref{fig:powerspectra}(c).  Gollub and
Benson~\cite{gollub:jfm_1980} and Curry {\it et. al.}~\cite{curry:jfm_1984} report $f_1/f_2$ to be
approximately 7/3 and 10/3 respectively for their phase-locked regime.  Our
$f_1/f_2 \approx 167/53$ is in general agreement with these earlier
results.

With a further increase in the bifurcation parameter,  the system
becomes chaotic at around $r=44.2$.  The route to chaos is similar
to that in Curry-Yorke model~(\S VIII.3 \cite{berge:book}) where
chaos appears after quasiperiodicity in $T^2$ (2-Torus) and phase locking. In
the bifurcation diagram (Fig.~\ref{bifurcation_model_a}), the
chaotic region is shown by colored pink patch labeled as `CH'. As
shown in Fig.~\ref{fig:powerspectra}(d) the power spectrum of the mode
$W_{101}$ is broad indicating chaotic nature of the attractor. At
around $r=45.9$, the size of the chaotic attractor suddenly
increases as a result of an `interior crisis'.  This feature is
illustrated in Fig.~\ref{fig:crisis} where we plot the phase space
projection on the $|W_{101}|-|W_{103}|$ plane  at $r=45.8$ and
$r=45.9$.

The chaotic state described above exists till $r \leq 46.2$ after
which we observe periodic solutions (the red curves in
Fig.~\ref{bifurcation_model_a}) that emerge from an inverse NS
bifurcation.  This inverse NS bifurcation is illustrated in
Fig.~\ref{floquet_plane}(c) wherein the largest Floquet multipliers
enters into the unit circle making the limit cycles stable. This
stable time-periodic orbit continues till $r=47.4$ at which point
another pair of Floquet multiplier again crosses the unit circle in
a forward NS bifurcation (see Fig.~\ref{floquet_plane}(d)) giving
rise to a quasi-periodic state.  In the periodic window the largest
Floquet multiplier is 1 as illustrated by the `BC' window in
Fig.~\ref{floquet}.   This quasi-periodic state subsequently becomes
chaotic at $r=48.4$ that continues for higher values of $r$. A
zoomed portion of this regime of $r$ is shown in the upper inset of
Fig.~\ref{bifurcation_model_a}.  We also note that the size of the
chaotic attractor is much larger than the QP attractor.  This
feature can be explained using `attractor-merging crisis' to be
discussed below.

\begin{figure}[t]
\begin{center}
\includegraphics[height=!,width=14cm]{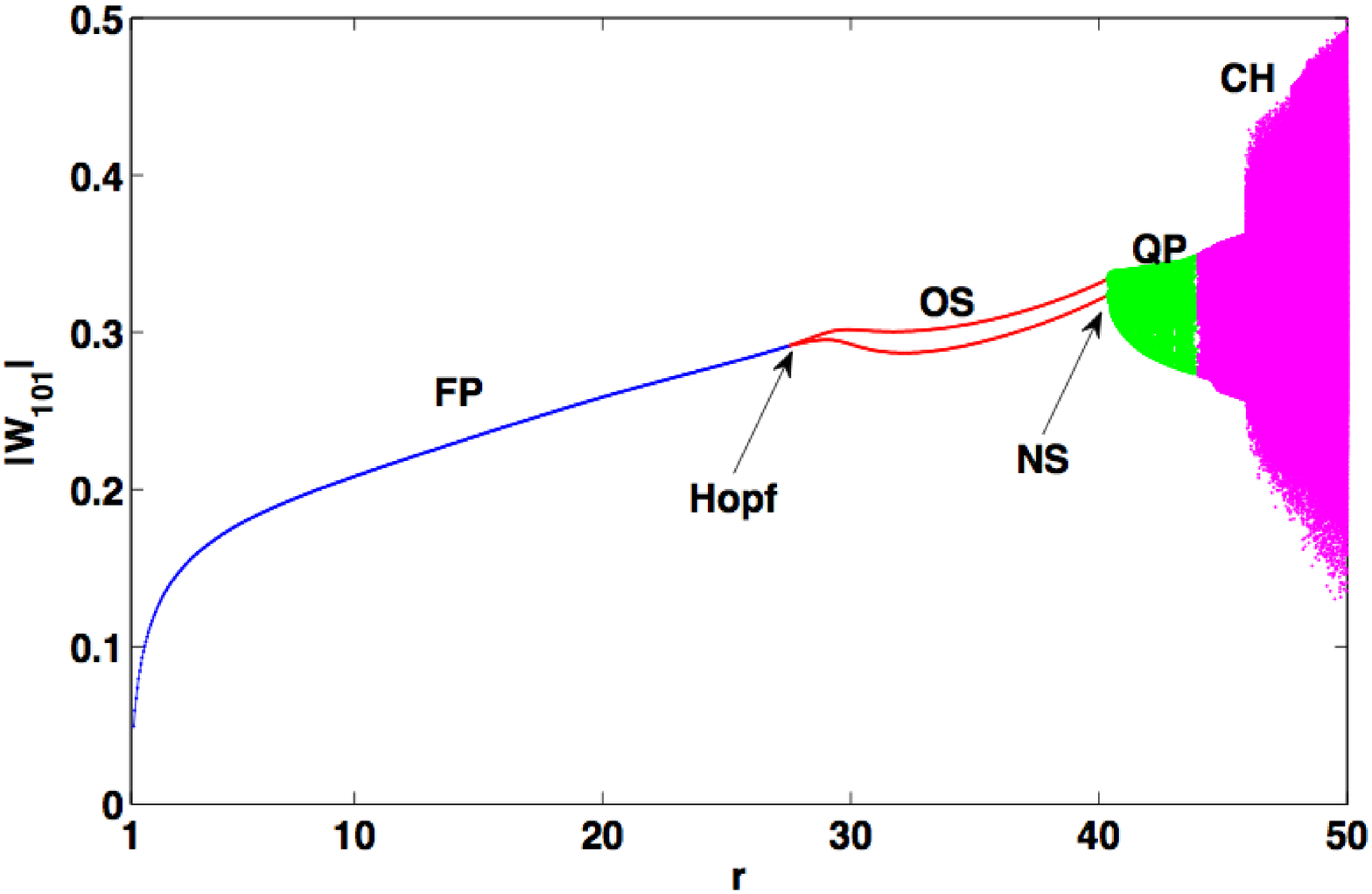}
\end{center}
\caption{Bifurcation diagram for  the low-dimensional model with initial condition different from the one used for generating Fig.~\ref{bifurcation_model_a}.
 `FP' stands for fixed point state (blue),
`OS' stands for oscillatory state (red), `QP' stands for
quasi-periodic state (green) and `CH' stands for chaotic state
(pink). `NS' stands for the Neimark-Sacker bifurcation point.  The present attractor and that of Fig.~\ref{bifurcation_model_a} differ for $r=46.2-48.4$. }
\label{bifurcation_model_b}
\end{figure}

We discussed earlier that in the band $r=46.2-48.4$ the
low-dimensional model has a periodic and a quasiperiodic attractor.
However, in the same band of $r$, a different set of initial
conditions yield another attractor which is chaotic (see
Fig.~\ref{bifurcation_model_b}).  These two attractors have been
shown in Fig.~\ref{fig:crisis2}(a), with the green region as the QP
attractor and the pink region as the chaotic attractor.  At $r=48.4$
these two attractors merge through `attractor-merging crisis' and
form a single large attractor shown in Fig.~\ref{fig:crisis2}(b).
The size of the resulting attractor is much larger that the original
QP attractor but similar to that of the chaotic attractor of
Fig.~\ref{bifurcation_model_b}.

\begin{figure}[t]
\begin{center}
\includegraphics[height=!,width=12cm]{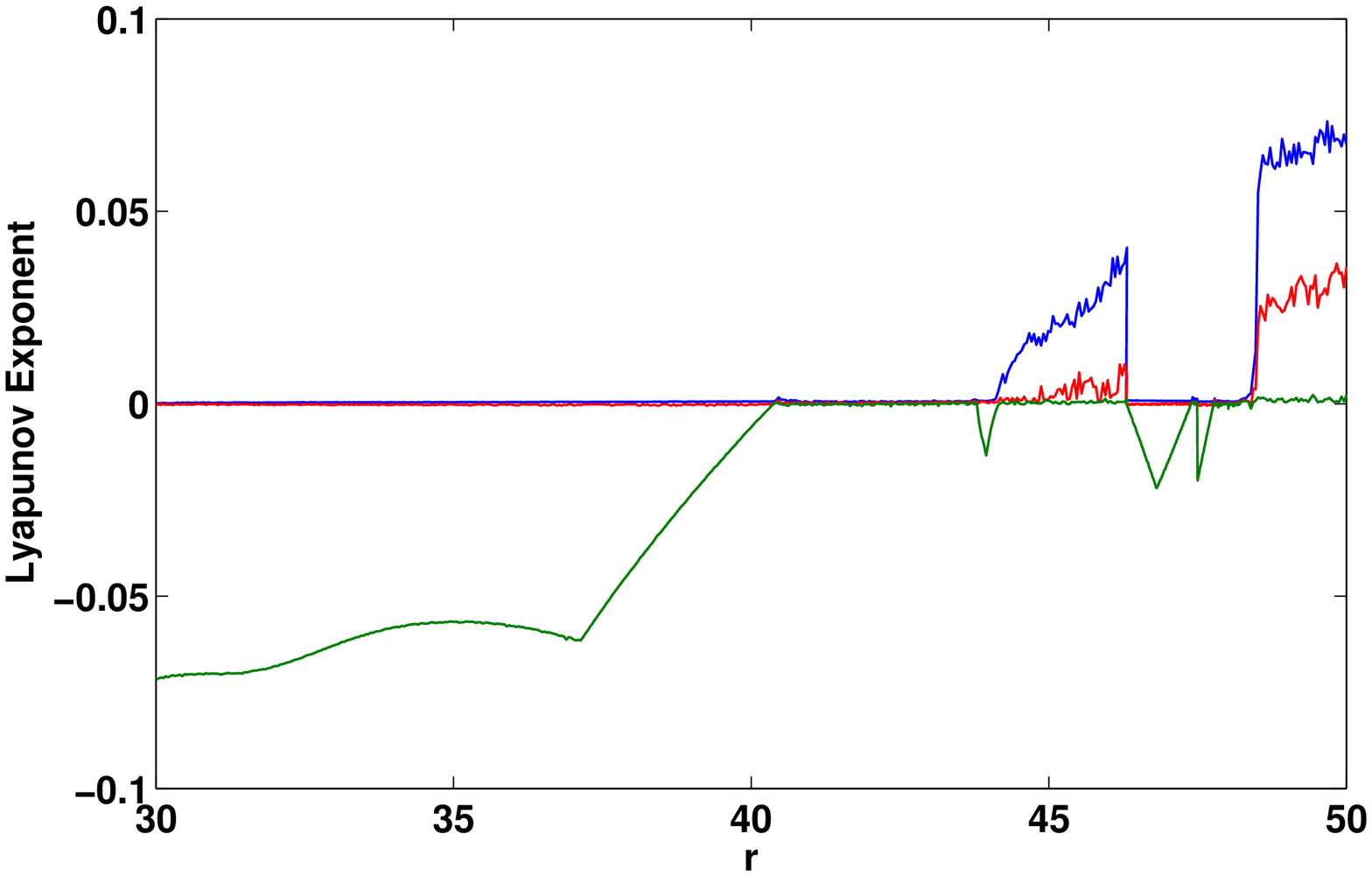}
\end{center}
\caption{Three largest Lyapunov exponents of  the low-dimensional model corresponding to the attractors in Fig.~\ref{bifurcation_model_a}.}
\label{lyapunov_exponent1}
\end{figure}

\begin{figure}[t]
\begin{center}
\includegraphics[height=!,width=12cm]{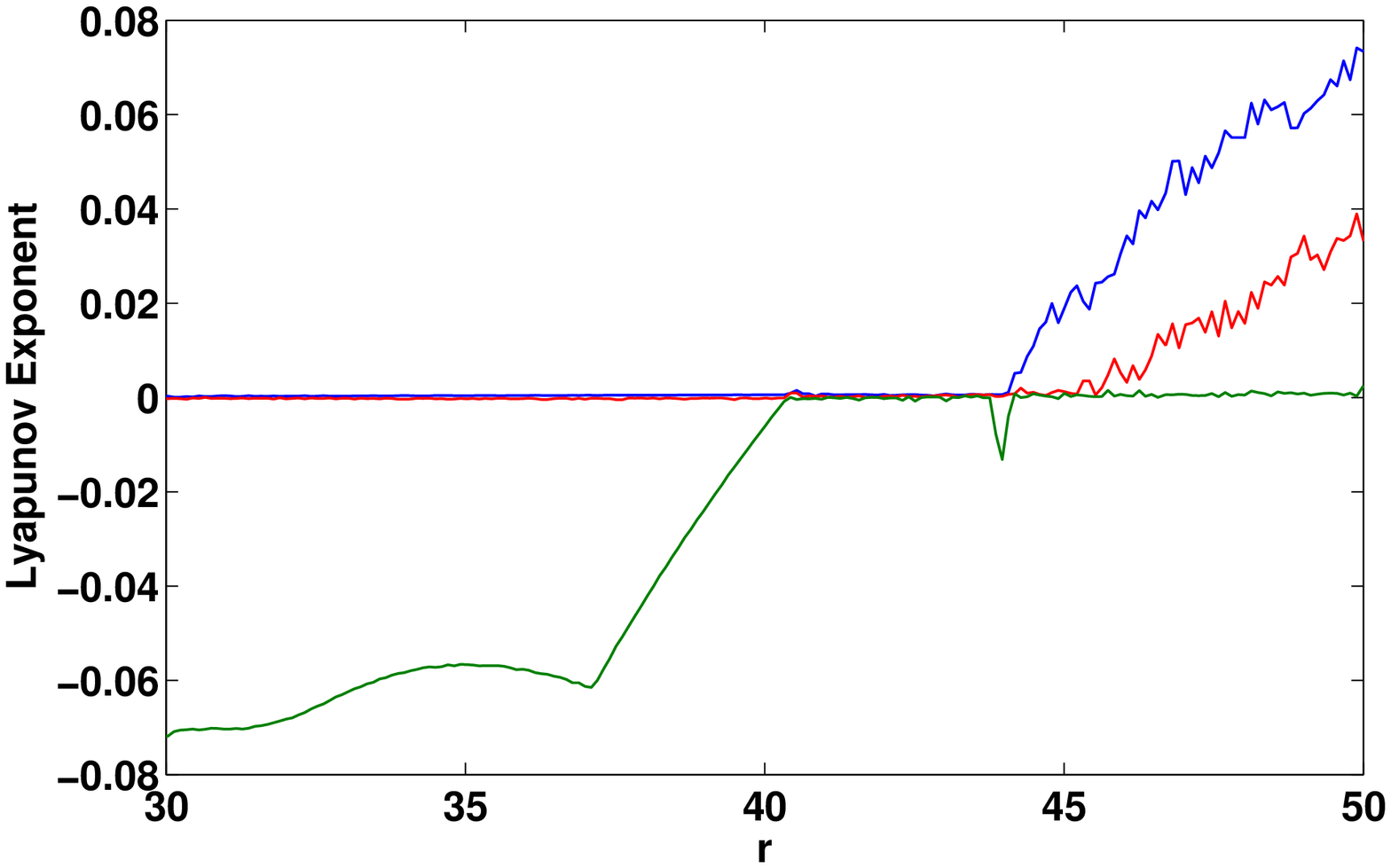}
\end{center}
\caption{Three largest Lyapunov exponents of the low-dimensional model corresponding to the attractors in Fig.~\ref{bifurcation_model_b}.}
 \label{lyapunov_exponent2}
\end{figure}

To ascertain the chaotic nature of the solutions obtained in our
low-dimensional system, we  compute the Lyapunov exponents
associated with the various solutions presented in
Figs.~\ref{bifurcation_model_a} and \ref{bifurcation_model_b}.  The
three largest Lyapunov exponents of our system corresponding to the
attractors in Figs.~\ref{bifurcation_model_a} and
\ref{bifurcation_model_b} are shown in
Figs.~\ref{lyapunov_exponent1} and \ref{lyapunov_exponent2}
respectively. There is at least one zero Lyapunov exponent throughout
the range consistent with the fact that our system is autonomous.
In the chaotic regions described earlier there are two positive Lyapunov exponents clearly distinct from
the zero exponent ascertaining the chaotic nature of the solutions.  The largest Lyapunov exponents in
Fig.~\ref{lyapunov_exponent1} is zero for $r = 46.2-48.4$ that
corresponds to the periodic and quasiperiodic window shown in
Fig.~\ref{bifurcation_model_a}.

\section{Discussions and Conclusions}

In this paper we present a bifurcation analysis of a 30-mode model
for the Prandtl number $P=6.8$ (a typical large-Prandtl number
fluid) and aspect ration $2\sqrt{2}$.    In our bifurcation analysis
we observe various patterns: steady rolls ($r=1-27.6$),
time-periodic rolls ($r=27.6-40.3$), quasiperiodicity
($r=40.3-43.8$), phase locking ($r=43.8-44.2$), and chaos
($r>44.2$).  The route to chaos is similar to that of
Curry-Yorke model~\cite{berge:book} where chaos occurs after
quasiperiodicity and phase locking.    Periodic and quasiperiodic
rolls reappear after the chaotic state in the range of $r =
46.2-47.4$ and $r=47.4-48.4$ respectively.  After the second
quasiperiodic window the system  becomes chaotic again through
`crisis'.   A distinct feature of our low-dimensional model is that
we track the fixed points, limit cycles, and chaotic attractors,
thus getting a detailed bifurcation picture for the range of $r$
under investigation.

The above features of our 30-mode model closely resemble some of the
past experimental results on large-Prandtl number convection namely
that of Gollub and Benson~\cite{gollub:jfm_1980} who observed chaos
in water for various Prandtl numbers and aspect ratios.  The route to
chaos in Gollub and Benson's experiment for $P=5$ and aspect ratio
of 3.5 is through quasiperiodicity and phase locking. In direct
numerical simulation of 3D RBC, Curry {\em et al.}~\cite{curry:jfm_1984} and
Yahata~\cite{Yahata:2000} observed similar transition to chaos for
$P=10$ and $P=5$ respectively.   Our low-dimensional model follow
the same route to chaos. The range of Rayleigh numbers for our
low-dimensional model is quite close to the Gollub and Benson's
experiments and Curry {\it et al.}'s and Yahata's DNS.  Thus our
low-dimensional model appears to capture the dynamics of 3D RBC
responsible for transition to chaos through quasiperiodicity and
phase locking.

Our low-dimensional model also exhibits coexistence of several
attractors. In the window of $r=46.2-48.4$, the system has periodic
and quasiperiodic attractor along with a chaotic attractor.
Coexistence of patterns and different attractors have been observed
earlier~\cite{Metcalfe}.  Another novel feature of our
low-dimensional model is that it reproduces reappearance of periodic
rolls after chaos, a feature observed in the 2D DNS of Curry {\it et
al.}~\cite{curry:jfm_1984} and Paul {\it et
al.}~\cite{paul:arxiv_2009}, albeit at a much
different $r$ value.   This feature appears to be due to the delay
of secondary instabilities in 2D DNS compared to 3D DNS.

Gollub and Benson~\cite{gollub:jfm_1980} also reported chaos in
their large-Prandl number RBC through period-doubling, generation of
three frequency (quasiperiodicity), and intermittency for different
sets of Prandtl numbers and aspect ratios.  Our preliminary
investigation for $P=10$ and aspect ratio of $2\sqrt{2}$ appears to
indicate intermittency, however, we need to study this phenomena
more carefully.  Some of the features reported by Gollub and Benson
could possibly be captured by our model by varying the aspect ratio, a
topic to be investigated in future.   Further work is required in construction
and analysis of more refined models for RBC. 

\section*{Acknowledgments}
We thank Krishna Kumar, Pankaj Mishra, and Pinaki Pal for useful
discussion.   Part of this work is supported by the grant of
Swarnajayanti fellowship to MKV by Department of Science and
Technology, India.  Our model is motivated by DNS runs performed in EKA of Computational Research Laboratory (CRL), Pune.  We thank CRL for providing us access to EKA.

\appendix
\section{Numerical search for periodic solutions of unknown periods through fixed points of a map}
\label{Poincare_map}

Consider a system of first order autonomous ODEs given by
\begin{equation}
\dot{\vec{x}} + \vec{f}(\alpha,\vec{x})  = 0\,, \label{arc_cont}
\end{equation}
where, $\vec{x} \in \mathbb{R}^n$, $\alpha$ is a parameter and
$\vec{f}(\alpha,\vec{x})$ is a known vector valued function and is
such that Eq.\ (\ref{arc_cont}) has a periodic solution.  For each
value of the parameter $\alpha$, we seek the initial conditions
$\vec{A}$ corresponding to a periodic solution and the time period
$T$ of the periodic solution. Hence, if we numerically integrate the
above equation, i.e., Eq.\ \ref{arc_cont}) with $\vec{x}(0)=\vec{A}$
till time $T$, we should have
\begin{equation}
\vec{x}(T)-\vec{A}\,=\,0\,. \label{newton}
\end{equation}
Equation (\ref{newton}) gives us $n$ algebraic equations. However,
we have $n+1$ unknowns, viz.\ the n components of $\vec{A}$ and the
time-period of the periodic solution $T$, and hence, we require one
more equation. This equation is obtained by putting a constraint
that the initial condition $\vec{A}$ correspond to the extremum of
one of the components, e.g., $x_1$. Accordingly, we will require
\begin{equation}
\dot {x}_1(T)\,=\, {f}_1(\alpha,\vec{x}(T))\,=  \, 0\,.
\label{newton1}
\end{equation}
Denoting \[ \vec{y} = \left(\begin{array}{c} \vec{A}\\ T
\end{array} \right), \] we can write Eqs.\ (\ref{newton}) and
(\ref{newton1}) as
\begin{equation}
\vec{g}(\vec{y})=0. \label{newton2}
\end{equation}
Equation (\ref{newton2}) can now be solved numerically using the
Newton-Raphson method. Note that evaluation of $g$ in the solution
procedure involves background numerical solution of the ODEs (Eq.\
(\ref{arc_cont})). This numerical integration can be avoided if an
analytical solution were available for the ODEs such that the map
$\vec{g}$ would be known analytically. However, the basic principle
of constructing the map $\vec{y}=\vec{g}(\vec{y})$ whose fixed
points gives us the appropriate initial conditions and the
time-period of the unknown periodic solution of the ODEs remains the
same.

\end{document}